\documentclass[aps,preprint,nofootinbib,tightenlines,superscriptaddress]{revtex4}
\usepackage{epsfig}
\usepackage{amsmath}
\usepackage{graphicx}
\newcommand{\beq}{\begin{equation}}
\newcommand{\eeq}{\end{equation}}
\newcommand{\bqa}{\begin{eqnarray}}
\newcommand{\eqa}{\end{eqnarray}}
\newcommand{\nnb}{\nonumber\\}

\begin{document}

\title{Low Transverse Momentum Heavy Quark Pair Production to
Probe Gluon Tomography}

\author{Ruilin Zhu}
\affiliation{School of Physics, University of Chinese Academy of
Sciences, Beijing, China} \affiliation{Nuclear Science Division,
Lawrence Berkeley National Laboratory, Berkeley, CA 94720, USA}
\author{Peng Sun}
\affiliation{Nuclear Science Division, Lawrence Berkeley National
Laboratory, Berkeley, CA 94720, USA}
\author{Feng Yuan}
\affiliation{Nuclear Science Division, Lawrence Berkeley National
Laboratory, Berkeley, CA 94720, USA}

%\date{\today}
\begin{abstract}
We derive the transverse momentum dependent (TMD) factorization for
heavy quark pair production in deep inelastic scattering, where the
total transverse momentum is much smaller than the invariant mass of
the pair.  The factorization is demonstrated at one-loop order, in
both Ji-Ma-Yuan and Collins-11 schemes for the TMD definitions, and
the hard factors are calculated accordingly. Our result provides a
solid theoretical foundation  for the phenomenological
investigations of the gluon TMDs in this process, and can be
extended to other similar hard processes, including dijet
(di-hadron) production in DIS.
%The phenomenological implication is also discussed.

%\item[Keywords]
% 12.38.Bx   Perturbative calculations
% 12.39.St  Factorization
% 13.85.Fb Inelastic scattering: two-particle final states
% 14.65.Dw  Charmed quarks
\end{abstract}

\maketitle

\section{Introduction}

Nucleon tomography in terms of various quark and gluon distribution
functions has attracted great attention from both experiment and
theory sides in recent years~\cite{Boer:2011fh}. Among these
concepts, the transverse momentum dependent (TMD) parton
distributions offer a clear picture of internal parton structure of
the nucleon, and most importantly, they can be measured in hard QCD
scattering processes. Together with the generalized parton
distributions (GPDs), they reveal the nucleon tomography in most
complete fashion, i.e., the three-dimension imaging of partons in
nucleon. Theory and experiment developments toward this tomography
has been one of focused topics in the planed electron-ion collider
(EIC)~\cite{Boer:2011fh}.  The TMD quark distributions can be
systematically investigated in the semi-inclusive hadron production
in deep inelastic scattering processes (SIDIS), and a great deal
have been learned from previous experiments, and will be mostly
covered in the future experiments, such as JLab 12 GeV upgrade and
the EIC. TMD quark distributions can also be studied in the
Drell-Yan lepton pair production in $pp$ collisions. On the other
hand, the gluon TMDs are not easy to access. This is because photon
does not couple to gluon at leading power in the hard scattering
processes. It has been suggested to probe the gluon TMDs through
two-particle productions in DIS or $pp$ collision processes, such as
heavy quark pair production in
DIS~\cite{Dominguez:2010xd,Boer:2010zf}, direct two photon
production in $pp$ collisions~\cite{Qiu:2011ai,Balazs:2007hr}. The
associated QCD dynamics for two-particle production has attracted
intensive theoretical studies in recent years, including the
factorization property for the hard processes and the universality
of the parton
distributions\cite{Collins:2007nk,Mulders:2011zt,Vogelsang:2007jk}.
It was generally believed that a TMD factorization shall apply to
two-particle production in DIS processes. However, it has never been
explicitly written down in a factorization form of TMD
distributions. In the following, we will, for the first time, derive
the TMD factorization formula for low transverse momentum heavy
quark pair production. This result shall provide a solid theoretical
foundation for phenomenological studies of gluon tomography in hard
process, and open a new window for QCD studies for various other
processes as well, such as dijet (di-hadron) production in DIS.

We focus on the heavy quark pair production in DIS
process\cite{Ellis:1988sb},
\begin{equation}
\gamma^*+p\to c\bar c[M_{c\bar c},p_\perp]+X \ ,
\end{equation}
 where the transverse momentum of the pair $p_\perp$
is much smaller than the invariant mass $M_{c\bar c}$, and we keep
the virtuality of the  photon the same order as $M_{c\bar
c}$~\footnote{This can be generalized to real photon case as well,
where a similar TMD factorization can be derived.}. In the following
calculations, we denote incoming photon momentum as $q$, $P_A$ for
nucleon (along +$\hat z$ direction), $k_1$ and $k_2$ for the heavy
quark and antiquark, respectively. We further introduce two
dimensionless vectors $n_c=k_1/m_c$ and $n_{\bar c}=k_2/m_c$ to
represent the directions of two final state particle: $n_c^2=n_{\bar
c}^2=1$. In addition, two light-like vectors are adopted:
$n=(1^-,0^+,0_\perp)$ and $\bar n=(0^-,1^+,0_\perp)$, where $\pm$ of
a momentum is defined as $P^\pm=(P^0\pm P^z)/\sqrt{2}$. Therefore,
$P_A$ is dominated by its plus component. In this kinematics, the
differential cross section is sensitive to the transverse momentum
dependence of the gluon in the nucleon, and we can formulate a TMD
factorization. Because the final state carries color, the naive TMD
factorization has to be modified. From our following calculations,
we find that an additional soft factor shall be included in the
factorization formula. Therefore, the differential cross section can
be written as
\begin{equation}
\frac{d\sigma(\gamma^*p\to c\bar c+X)}{d^2p_\perp dM_{c\bar c}^2
d\cos\theta}=
\sigma_0\int\frac{d^2b_\perp}{(2\pi)^2}e^{ip_\perp\cdot b_\perp}
W_{c\bar c}(x, b_\perp) \ ,
\end{equation}
where $\sigma_0$ represents the leading Born diagrams contribution,
\bqa
 \sigma_0&=&\frac{\pi^2 M_{c\bar
c}^2 N_c C_F\alpha_s\alpha e_c^2\beta}{
(N_c^2-1)(S+Q^2)\widetilde{Q}^4}\{\frac{t_1}{u_1}+\frac{u_1}{t_1}+\frac{4m_c^2\widetilde{Q}^2}{t_1
u_1}(1-\frac{m_c^2\widetilde{Q}^2}{t_1u_1})\nnb&&-\frac{2 Q^2 }{t_1
u_1}(\widetilde{Q}^2+ \frac{2t_1u_1-\widetilde{Q}^4}{t_1
u_1}m_c^2)-\frac{2 Q^4}{t_1 u_1}\}, \eqa here  $Q^2=-q^2$ with $q$
the momentum of the virtual photon, $\beta=\sqrt{1-4m_c^2/M_{c\bar
c}^2}$, and the variable $\widetilde{Q}$ is defined as
$\widetilde{Q}^2=M_{c\bar c}^2+Q^2$. Then the longitudinal-momentum
fraction  for the incident gluon from nucleon $x$ can be written as
$x=\widetilde{Q}^2/(S+Q^2)$ with the photon-hadron center-of-mass
energy $S=(q+P_A)^2$. In the above expression,
 $\theta$ is
the scattering angle between final heavy quark and nucleon in the
photon-hadron center-of-mass frame. $t_1=(k_1-xP_A)^2-m_c^2$, and
$u_1=(k_2-xP_A)^2-m_c^2$.  In the impact parameter $b_\perp$-space,
the TMD factorization for $W_{c\bar c}$ can be written as
\begin{equation}
W_{c\bar c}(x,b_\perp)=H(\widetilde{Q},\mu)
xg(x,b_\perp,\widetilde{Q},\mu) \overline{S}(b_\perp,\mu)
\label{factorization}\ ,
\end{equation}
where the hard factor $H$, the soft factor $\overline{S}$, and the
TMD gluon distribution all depend on the factorization scale $\mu$.
Schematically, this factorization can be viewed as in Fig.~1, where
the photon scatters off gluon to produce the heavy quark pair (hard
factor), and the transverse momentum of the final state comes from
the gluon distribution (lower part) and the soft gluon radiation
from the final state. Compared to the SIDIS, we will find that in
the above factorization we have the soft function $\overline{S}$
instead of the TMD fragmentation function. Naive TMD definitions for
the gluon distribution and soft function $\overline{S}$  contain the
light-cone singularities, and a regulation introduces the scheme
dependence. In the following, we will show calculations in two
schemes: Ji-Ma-Yuan scheme~\cite{Ji:2004wu} and Collins-11
scheme~\cite{Collins}. We first derive the soft factor in this
process, and calculate the associated hard factors and the gluon
TMDs. In the end, we will derive the resummation formalism. We will
find that the final results do not depend on the schemes.

\begin{figure}[t]
\begin{center}
\includegraphics[width=0.5\textwidth]{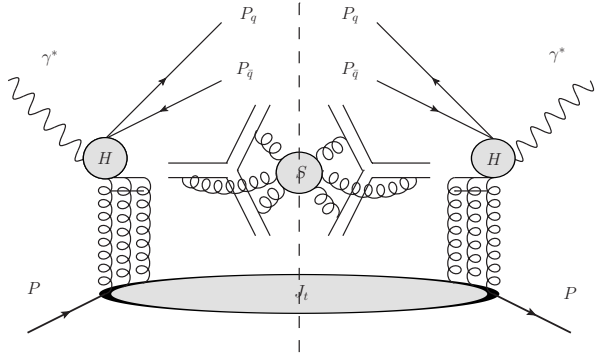}
\end{center}
\vskip -0.7cm \caption{TMD factorization for heavy quark pair
production in DIS process.}
\end{figure}

\section{Soft Gluon Radiation}

A key point to demonstrate the QCD factorization is to show that the
leading power gluon radiation can be included into the various
factors in the factorization formula. In this analysis, a power
counting method is crucial to achieve the final factorization
result. For example, the gluon radiation associated with the
incoming gluon contributes to the collinear and soft gluon part,
which can be absorbed into the gluon distribution and soft factor in
the  factorization formula. This part of contribution is similar to
those in the SIDIS and low transverse momentum Drell-Yan lepton pair
production in $pp$ collisions. Now, we turn to the final state
radiation. Because of heavy quark mass, soft radiation from the
quark pair contributes to the leading power of the differential
cross section.  The resummation of all order soft gluon radiation
associated with heavy quark moving in $n_c$-direction can be
summarized into a Wilson line in that direction. This has also been
applied to formulate the threshold resummation in heavy quark pair
production\cite{Kidonakis:1997gm,Botts:1989kf}. By applying this
technique, we can summarize the gluon radiation from the initial
state gluon and the final state heavy quark pair as three kinds of
Wilson lines. For finial-state quark and anti-quark, we have \bqa
{\cal L}_{n_c}(\xi)&=&P\exp\left(-ig\int_{0}^{+\infty}
    d\lambda n_c\cdot A(\lambda n_c + \xi) \right),\nnb
{\cal L}_{n_{\bar{c}}}(\xi)&=&P\exp\left(ig\int_{0}^{+\infty}
    d\lambda n_{\bar{c}}\cdot A(\lambda n_{\bar{c}} + \xi) \right),
\eqa which are in the fundamental representation, $A^\mu= A_c^\mu
T^c$. While for the soft gluon radiation from the incoming gluon, we
have
\begin{equation}
{\cal L}_{\bar{v}}(\xi)=P\exp\left(ig\int_{0}^{+\infty}
    d\lambda \bar{v}\cdot A(\lambda\bar{ v} + \xi) \right),
\end{equation}
which is in the adjoint representation, $A^\mu=-if_{abc}  A_b^\mu $.
The vector $\bar{v}$ is along the momentum $P_A$. The soft function
in the factorization formula Eq.~(\ref{factorization}) contains
contributions from the above three Wilson lines. For the process of
 Eq.~(1), there is only one color base combination between them, and the soft
function can be written as,

\begin{figure}[t]
\begin{center}
\includegraphics[width=0.68\textwidth]{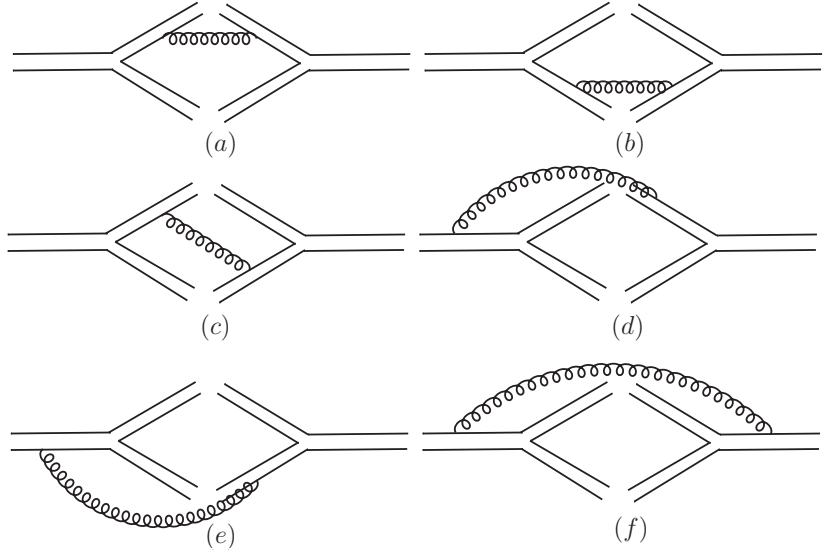}
\end{center}
\vskip -0.7cm \caption{Real corrections to the soft function.}
\end{figure}

%\begin{equation}
%S^{(\bar v,n_t,n_{\bar t})}(b,\mu,\rho)=\frac{\langle 0|{\cal L}_{\bar
%vca'}^\dagger(b_\perp;\infty) \textmd{Tr}\left[ {\cal
%L}_{n_c}^\dagger(b_\perp;\infty) \textmd{T}^{a'} {\cal
%L}_{n_{\bar{c}}}^\dagger(\infty;b_\perp) {\cal
%L}_{n_{\bar{c}}}(0;\infty)\textmd{T}^{a}{\cal L}_{n_c}(\infty;0)
%\right] {\cal L}_{\bar
%vac}(\infty;0)|0\rangle}{\textmd{Tr}[\textmd{T}^d \textmd{T}^d]}\, ,
%\label{softdef}
%\end{equation}

\begin{equation}
\overline{S}^{}(b_\perp,\mu,\rho)=\frac{\int_0^\pi
\frac{(\sin\phi)^{-2\epsilon}}{a_1}d\phi\;\langle 0|{\cal L}_{\bar
vca'}^\dagger(b_\perp) \textmd{Tr}\left[ {\cal
L}_{n_c}^\dagger(b_\perp) \textmd{T}^{a'} {\cal
L}_{n_{\bar{c}}}^\dagger(b_\perp) {\cal
L}_{n_{\bar{c}}}(0)\textmd{T}^{a}{\cal L}_{n_c}(0) \right] {\cal
L}_{\bar vac}(0)|0\rangle}{\textmd{Tr}[\textmd{T}^d \textmd{T}^d]}\,
, \label{softdef}
\end{equation}
with
$a_1=\frac{\sqrt{\pi}\Gamma(\frac{1}{2}-\epsilon)}{\Gamma(1-\epsilon)}$
in the $(4-2\epsilon)$ dimension. In the above definition, we have
integrated out the azimuthal angle $\phi$ between $n_{c\perp}$ and
$p_\perp$. Intuitively, $\bar v$ shall be chosen along the momentum
direction of the nucleon $P_A$ with $\bar v=\bar n$. However, this
definition contains a light-cone singularity~\footnote{There is no
light-cone singularity associated with  ${\cal L}_{n_c,n_{\bar c}}$
because of $n_{c,\bar c}^2\neq 0$.}. Regulating this singularity
introduces the scheme-dependence. In our calculations, we follow two
different schemes: Ji-Ma-Yuan~\cite{Ji:2004wu} and
Collins-11~\cite{Collins}. We will show that the factorization works
for both schemes, and the hard factors can be calculated
accordingly.

In the Ji-Ma-Yuan scheme, the light-cone gauge link  is chosen
slightly off-light-cone: $\bar v^2\neq 0$ with $\bar v^+\gg \bar
v^-$. In Fig.~2, we plot the real gluon radiation contribution to
the soft factor. Similarly, we have virtual diagrams. Adding them
together, we find that, at one-loop order,
\begin{eqnarray}
\overline{S}_{\rm JMY}^{(1)}(b_\perp,\mu,\rho) &=&
\frac{\alpha_s}{2\pi}\{{C_A}\ln{\frac{c_0^2}{b_\perp^2\mu^2}}\;(B_{final}+\ln
\rho^2+\ln\frac{\widetilde{Q}^2}{\zeta^2}-1)+C_{final}\} \
,\label{softf}
\end{eqnarray}
where $c_0=2e^{-\gamma_E}$, $\rho^2=(2v\cdot \bar v)^2/v^2\bar v^2$,
$\zeta^2=x^2(2v\cdot P_A)^2/v^2$, and $v$ is another non-light-like
vector $v^2\neq 0$ with $v^-\gg v^+$. B and C functions are defined
as
\begin{eqnarray}B_{final}&=&\frac{1}{2N_c^2}\frac{1+\beta^2}{\beta}\ln\frac{1-\beta}{1+\beta}
-2\frac{C_F}{N_c} +\ln{\frac{t_1u_1}{\widetilde{Q}^2\;m_c^2} }\ ,\nonumber\\
C_{final}&=&\frac{1}{2N_c}f_{c\bar{c}}
+2C_F\ln{\frac{t_1u_1}{M_{c\bar c}^2\;m_c^2} } -{C_A}\;\mathrm{Li}_2
(1-\frac{t_1u_1}{M_{c\bar c}^2\;m_c^2}) \ .
\end{eqnarray}
The large $N_c$ suppressed term $f_{c\bar{c}}$ can be further
decomposed into
\begin{eqnarray}f_{c\bar{c}}&=& -\frac{\left(\beta ^2+1\right)}{2 \beta }f^a_{c\bar{c}}+\frac{2 \left(\beta ^2+1\right) (1-\beta  \cos \theta )}{1-\beta ^2}f^b_{c\bar{c}}\
,
\end{eqnarray}
where the factor $f^a_{c\bar{c}}$ can be written as
\begin{eqnarray}f^a_{c\bar{c}}&=&\left(\ln
   \frac{b_1}{b_4}\right)^2 -\left(\ln \frac{b_3}{b_2}\right)^2+2 \ln \frac{b_3 b_4}{b_1 b_2} \ln \frac{b_1 \cot ^2\left(\frac{\theta
   }{2}\right)}{b_2}\nnb&&+2
   \left(\text{Li}_2(\frac
   {b_2}{b_4})+\text{Li}_2(\frac{b_4}{b_1})-\text{Li}_2(\frac{b_2}{b_3})-\text{Li}_2(\frac{b_3}{b_1})\right)\
   ,
\end{eqnarray}
with $b_1=1+\beta$, $b_2=1-\beta$, $b_3=1+\beta\cos \theta$,
$b_4=1-\beta\cos \theta$. The analytic expression of
$f^b_{c\bar{c}}$ is complicated to present in the paper, so we will
reserve the integral
\begin{eqnarray}f^b_{c\bar{c}}&=&\int_0^1 \frac{\ln (a y^2+1)}{-a y^2+y (a+c-1)+1}\
   ,
\end{eqnarray}
with $a=\beta^2\sin^2\theta/(b_1 b_2)$ and $c=b_4^2/(b_1 b_2)$.

 Similarly, the non-light-like vector $v$ defined above was introduced to regulate the
light-cone singularity in the TMD gluon distribution in Ji-Ma-Yuan
scheme, for which we have~\cite{Ji:2005nu}
\begin{eqnarray}
xg^{unsub.}(x,k_\perp,\mu,\zeta,\rho)&=&\int\frac{d\xi^-d^2\xi_\perp}{P^+(2\pi)^3}
    e^{-ixP^+\xi^-+i\vec{k}_\perp\cdot \vec\xi_\perp}\nonumber\\
    &&\times
    \left\langle P|{F_a^+}_\mu(\xi^-,\xi_\perp)
{\cal L}^\dagger_{vab}(\xi^-,\xi_\perp) {\cal L}_{vbc}(0,0_\perp)
F_c^{\mu+}(0)|P \right\rangle\ ,\label{gpdfu}
\end{eqnarray}
where the associated gauge link is in adjoint representation. The
above gluon distribution contains not only collinear gluon
contribution, but also the soft gluon contribution, which is defined
as
%\begin{equation}
%S^{v,\bar v}(b_\perp)={\langle 0|{\cal L}_{\bar
%vcb'}^\dagger(b_\perp;\infty) {\cal
%L}_{vb'a}^\dagger(\infty;b_\perp){\cal L}_{vab}(0;\infty){\cal
%L}_{\bar vbc}(\infty;0)  |0\rangle/(N_c^2-1)   }\, . \label{softg}
%\end{equation}
\begin{equation}
S^{v,\bar v}(b_\perp)={\langle 0|{\cal L}_{\bar
vcb'}^\dagger(b_\perp) {\cal
L}_{vb'a}^\dagger(b_\perp){\cal L}_{vab}(0){\cal
L}_{\bar vbc}(0)  |0\rangle/(N_c^2-1)   }\, . \label{softg}
\end{equation}
Therefore the gluon distribution after subtraction is defined as
\begin{equation}
xg_{\rm
JMY}(x,b_\perp,\mu,\zeta,\rho)=xg^{unsub.}(x,b_\perp,\mu,\zeta,\rho)/S^{v,\bar
v}(b_\perp) \ ,
\end{equation}
which will enter into the factorization formula in
Eq.~(\ref{factorization}).

To demonstrate the TMD factorization at one-loop order, we calculate
the differential cross section on an on-shell gluon target. In the
perturbative calculations, we take the leading power contribution in
the limit $\widetilde{Q}\gg q_\perp$. It can be shown, in this
limit, diagram by diagram that they can be factorized into the TMD
gluon distribution, soft factor and the hard factor as in the
factorization formula. In particular, the one-loop result for the
TMD gluon distribution has been calculated in Ref.~\cite{Ji:2005nu}
in Ji-Ma-Yuan scheme. By subtracting the gluon distribution, we find
that the hard factor can be written as
\begin{eqnarray}
H_{\rm JMY}^{(1)}(\mu,\rho)&=&\frac{\alpha_s\,C_A}{\pi} \left\{
(\beta_0-\frac{B_{final}}{2}+\frac{\ln\rho}{2}-\frac{3}{4})\ln\frac{\widetilde{Q}^2}{\mu^2}
+\frac{\ln^2\rho}{4}-\frac{3}{4}\ln\rho
    +\frac{\pi^2}{6}+\frac{7}{4}+B^{V,g\gamma}_f\right\}\label{hardk}\
    ,\nnb
\end{eqnarray}
where we have taken $\zeta^2=\rho \widetilde{Q}^2$ for convenience,
and $B^{V,g\gamma}_f$ comes from the finite contribution of the
virtual diagrams.

%, and subtract the TMD gluon
%distribution and soft factor to obtain the hard factor.
%After performing the Fourier transformation into the impact
%parameter b space, we gain
%\begin{eqnarray}
%&&W^{(1)}(x,b,M^2,\mu)=\sigma_{g\gamma}^{(o)}\frac{\alpha_sN_c}{\pi}\left\{
%P_{gg}(x)\left(-(\frac{1}{\epsilon_{\rm IR}}+\ln4\pi-\gamma_E)-\ln\frac{b^2\mu^2}{4}e^{2\gamma_E}\right)\right.\nonumber\\
%    &&+\left.\delta(x-1)
%\left[(\beta_0-\frac{B_{final}}{2})\ln\left(\frac{M^2b^2}{b_0^2}\right)-\frac{1}{4}\ln^2\left(\frac{M^2b^2}{b_0^2}\right)
%    -\frac{\pi^2}{12}+
%    \frac{B^{V,g\gamma}_{f,b}}{2}\right]\right\}\ ,
%\end{eqnarray}

%\begin{equation}
%{\cal P}_{gg}(x)=
%    \frac{x}{(1-x)_+}+\frac{1-x}{x}+x(1-x)
%    +\delta(x-1)\beta_0\ ,
%\end{equation}
%with $\beta_0=(11N_c-2n_f)/(12N_c)$.

\section{Collins-11 Scheme}

Subtraction of the light-cone singularity is essential to establish
the TMD factorization. Collins introduced a subtraction scheme where
the parton distribution and soft factor do not contain light-cone
singularity~\cite{Collins}. According to this new scheme, the TMD
gluon is defined as
\begin{equation}
xg_{\rm
JCC}(x,b_\perp,\mu,\zeta_c)=xg^{unsub.}(x,b_\perp)\sqrt{\frac{S^{\bar
n,v}(b_\perp)}{S^{n,\bar n}(b_\perp)S^{n,v}(b_\perp)}} \ ,
\end{equation}
where $\zeta_c^2=x^2(2v\cdot P_A)^2/v^2=2(xP_A^+)^2e^{-2y_n}$ with
$y_n$ the rapidity cut-off in Collins-11 scheme. Calculating up to
one-loop order, we have
\begin{eqnarray}
xg_{\rm JCC}^{(1)}(x,b_\perp,\mu,\zeta_c)&=&
\frac{\alpha_s}{2\pi}C_A\left\{2\left(-\frac{1}{\epsilon}+\ln\frac{c_0^2}{b_\perp^2\bar\mu^2}\right){\cal
P}_{g\to g}(x)
\right.\nonumber\\
&&\left.+\delta(1-x)\left[2\beta_0\ln\frac{b_\perp^2\mu^2}{c_0^2}
+\frac{1}{2}\left(\ln\frac{\zeta_c^2}{\mu^2}\right)^2-\frac{1}{2}\left(\ln\frac{\zeta_c^2b_\perp^2}{c_0^2}\right)^2
\right]\right\} \ ,
\end{eqnarray}
where the gluon splitting kernel ${\cal P}_{gg}(x)=\frac{x}{(1-x)_+}+\frac{1-x}{x}+x(1-x)+\delta(x-1)\beta_0$
with $\beta_0=(11N_c-2n_f)/(12N_c)$.

The soft factor can be defined similarly,
\begin{equation}
\overline{S}_{\rm JCC}(b_\perp,\mu)=\overline{S}^{(\bar n,
n_c,n_{\bar c})}(b_\perp)\sqrt{\frac{S^{n,\bar
v}(b_\perp)}{S^{n,\bar n}(b_\perp)S^{\bar n,\bar v}(b_\perp)}} \ ,
\end{equation}
which turns out to be
\begin{equation}
\overline{S}_{\rm
JCC}^{(1)}(b_\perp,\mu)=\frac{\alpha_s}{2\pi}\{{C_A}\ln{\frac{c_0^2}{b_\perp^2\mu^2}}(B_{final}+\ln\frac{\widetilde{Q}^2}{\zeta_c^2})+C_{final}\}
\ ,
\end{equation}
at one-loop order.
From the above results, we immediately obtain the hard factor as
\begin{equation}
H^{(1)}_{\rm JCC}(\mu)=\frac{\alpha_s\,C_A}{\pi} \left\{
(\beta_0-\frac{B_{final}}{2})\ln\frac{\widetilde{Q}^2}{\mu^2}
-\frac{1}{4}\ln^2\frac{\widetilde{Q}}{\mu}-\frac{\pi^2}{12}+B^{V,g\gamma}_f\right\}\
,
\end{equation}
where $\zeta_c$ has been chosen as $\widetilde{Q}$ in the above
calculation.
\section{Resummation}

The large logarithms in the fixed order perturbative calculations as
we have shown above can be resummed by applying the
Collins-Soper-Sterman resummation~\cite{Collins:1984kg}. In
particular, in this case, we can derive the energy evolution
equation for the TMD gluon distribution, and the renormalization
group equation for the soft and hard factors. By solving these
equations, we resum the large logarithms. The final expression for
$W(b_\perp,\widetilde{Q})$ can be written as
\begin{eqnarray}
W(x,b_\perp,\widetilde{Q}^2)
&=&g(x,b_\perp,\widetilde{Q}_0,\widetilde{Q}_0)
\overline{S}(b_\perp,\widetilde{Q}_0)H(\widetilde{Q},\widetilde{Q})e^{-{\cal
S}_{sud}(\widetilde{Q},\widetilde{Q}_0)} \ ,
\end{eqnarray}
where $\widetilde{Q}_0$ is chosen such that the intrinsic TMD gluon
distribution at the input scale $\widetilde{Q}_0$. All the large
logarithms is included in the Sudakov form factor,
\begin{equation}
S_{sud}= -\int^{\widetilde{Q}}_{\widetilde{Q}_0}\frac{d\mu
}{\mu}\left(\ln{\frac{\widetilde{Q}}{\mu}}\gamma_K(\mu)
-\gamma_S(\mu,1)+\frac{\alpha_sC_A}{\pi}
(1-2\beta_0-\ln\frac{\widetilde{Q}_0^2b_\perp^2}{c_0^2})\right) \ ,
\end{equation}
where \bqa \gamma_K(\mu)&=&\frac{2\alpha_s(\mu)
C_A}{\pi}\,,\nnb\gamma_S(\mu,\rho)&=&-\frac{\alpha_s(\mu)
C_A}{\pi}(B_{final}+\ln\rho-1)\,. \eqa We notice that the
$\rho$-dependence cancels out in the above Sudakov form factor, as
well as in the final expression in the resummed form. In particular,
in the Sudakov factor $S_{sud}$, $\gamma_S$ takes value at $\rho=1$
after the resummation. We would like to emphasize that the final
results agree with each other between the  Ji-Ma-Yuan and Collins-11
schemes of the TMD definitions.

The above factorization results can be carried out for the
spin-dependent observables in this process as well, in particular,
for the  single transverse spin asymmetry from the gluon Sivers
function, which will have a similar factorization formula with the
soft function defined above. The associated single spin asymmetry
can be written as $S_\perp\times p_\perp$ where $S_\perp$ is the
transverse spin vector and $p_\perp$ defined above as the total
transverse momentum of the heavy quark pair. This tells that the
heavy quark pair production in DIS at the planed EIC will provide
important information on the gluon Sivers
function~\cite{Boer:2011fh}.

In addition, our results can also be extended to the linearly
polarized gluon distribution contribution in the above
process~\cite{Boer:2010zf}. However, because this term is
proportional to $\cos 2\phi$ where $\phi$ is the azimuthal angle
between $k_{1\perp}$ and $p_\perp$, the soft function
${\overline{S}}$ of Eq.~(\ref{softdef}) will have to be modified to
include explicit dependence on $\phi$. Nevertheless, a TMD
factorization can be formulated for this case as well.

\section{Conclusion}

In this paper, by studying one-loop correction to heavy quark pair
production at low transverse momentum, we have derived the
associated TMD factorization formalism in DIS process. The
light-cone singularity regulation in the TMD and soft factor has
been performed in both Ji-Ma-Yuan and Collins-11 schemes, and the
hard factors were calculated accordingly. The final results have
been shown to be independent of the regulation schemes. The
factorization result of this paper should provide an important step
further to investigate the gluon tomography in nucleon in hard
processes. A number of extensions shall follow, including QCD
factorization studies for di-jet (di-hadron) production in DIS,
heavy quark pair and di-jet production in $pp$ collisions. In
particular, a recent calculation of $t\bar{t}$ pair production in
$pp$ collision has been analyzed in the soft-collinear-effective
theory~\cite{Zhu:2012ts}. A detailed comparison with the current
calculation showed that they are consistent with each other. We will
address these issues in future publications.

\section*{Acknowledgements }
We thank L.~Yang for discussions concerning the results in
Ref.~\cite{Zhu:2012ts}. R.~Z. thanks Prof. C.~F.~Qiao for discussions, and he
is partially supported by China Scholarship Council. This
work was partially supported by the U. S. Department of Energy via
grant DE-AC02-05CH11231.

\end{document}